\documentclass[aps,twocolumn,superscriptaddress,nofootinbib,longbibliography,notitlepage,floatfix]{revtex4-2}
\usepackage{graphicx,outlines,subfigure}
\usepackage{physics,graphicx}
\usepackage{tabularx}
\usepackage{amsmath,mathtools}
\usepackage{amstext,cuted}
\usepackage{amssymb}
\usepackage{xfrac}
\usepackage[colorlinks,citecolor=magenta]{hyperref}
\usepackage{graphicx,makecell}
\usepackage{amsmath}
\usepackage{amstext}
\usepackage{amssymb}
\usepackage{longtable,booktabs}
\usepackage{hyperref}
\usepackage{url}
\usepackage{subfigure}
\usepackage{dsfont}
\usepackage{booktabs}
\usepackage{amsbsy}
\usepackage{dcolumn}
\usepackage{amsthm}
\usepackage{times}
\usepackage{bm}
\usepackage{esint}
\usepackage{multirow}
\usepackage{hyperref}
\usepackage{cleveref}
\usepackage{amsmath}
\usepackage{amssymb}
\usepackage{mathrsfs}
\usepackage{amsbsy}
\usepackage{dcolumn}
\usepackage{bm}
\usepackage{multirow}
\usepackage{color}
\usepackage{extarrows}
\usepackage{datetime}
\usepackage{comment}
\usepackage[super]{nth}
\usepackage{tikz}
\usetikzlibrary{arrows.meta}
\hypersetup{
	colorlinks=true,
	linkcolor=blue,
	filecolor=blue,
	urlcolor=blue,
}

\usepackage{datetime}
\usepackage{comment}
\usepackage{lineno}
\graphicspath{{Pictures/}}
\usepackage{xcolor}

\usepackage{scalerel}
\usepackage{tikz}
\usetikzlibrary{svg.path}

\definecolor{orcidlogocol}{HTML}{A6CE39}
\tikzset{
  orcidlogo/.pic={
    \fill[orcidlogocol] svg{M256,128c0,70.7-57.3,128-128,128C57.3,256,0,198.7,0,128C0,57.3,57.3,0,128,0C198.7,0,256,57.3,256,128z};
    \fill[white] svg{M86.3,186.2H70.9V79.1h15.4v48.4V186.2z}
                 svg{M108.9,79.1h41.6c39.6,0,57,28.3,57,53.6c0,27.5-21.5,53.6-56.8,53.6h-41.8V79.1z M124.3,172.4h24.5c34.9,0,42.9-26.5,42.9-39.7c0-21.5-13.7-39.7-43.7-39.7h-23.7V172.4z}
                 svg{M88.7,56.8c0,5.5-4.5,10.1-10.1,10.1c-5.6,0-10.1-4.6-10.1-10.1c0-5.6,4.5-10.1,10.1-10.1C84.2,46.7,88.7,51.3,88.7,56.8z};
  }
}

\newcommand\orcidicon[1]{\href{https://orcid.org/#1}{\mbox{\scalerel*{
\begin{tikzpicture}[yscale=-1,transform shape]
\pic{orcidlogo};
\end{tikzpicture}
}{|}}}}

\begin{document}
\title{Quantum entanglement and non-Hermiticity in free-fermion systems}

\author{Li-Mei Chen}
\thanks{These authors contributed equally to this work.}

	\affiliation{Guangdong Provincial Key Laboratory of Magnetoelectric Physics and Devices, State Key Laboratory of Optoelectronic Materials and Technologies,
and School of Physics, Sun Yat-sen University, Guangzhou, 510275, China}
\author{Yao Zhou}
\thanks{These authors contributed equally to this work.}
	\affiliation{Guangdong Provincial Key Laboratory of Magnetoelectric Physics and Devices, State Key Laboratory of Optoelectronic Materials and Technologies,
and School of Physics, Sun Yat-sen University, Guangzhou, 510275, China}
\author{Shuai A. Chen}\thanks{These authors contributed equally to this work.}
\affiliation{Max Planck Institute for the Physics of Complex Systems, N\"{o}thnitzer Stra{\ss}e 38, Dresden 01187, Germany}
\affiliation{Department of Physics, The Hong Kong University of Science and Technology, Hong Kong SAR, China}
\author{Peng Ye\orcidicon{0000-0002-6251-677X}}
\email{yepeng5@mail.sysu.edu.cn}
	\affiliation{Guangdong Provincial Key Laboratory of Magnetoelectric Physics and Devices, State Key Laboratory of Optoelectronic Materials and Technologies,
and School of Physics, Sun Yat-sen University, Guangzhou, 510275, China}

\date{\today}

\begin{abstract}
This topical review article reports rapid progress on the generalization and application of entanglement in non-Hermitian free-fermion quantum systems. We begin by examining the realization of non-Hermitian quantum systems through the Lindblad master equation, alongside a review of typical non-Hermitian free-fermion systems that exhibit unique features. A pedagogical discussion is provided on the relationship between entanglement quantities and the correlation matrix in Hermitian systems. Building on this foundation, we focus on how entanglement concepts are extended to non-Hermitian systems from their Hermitian free-fermion counterparts, with a review of the general properties that emerge. Finally, we highlight various concrete studies, demonstrating that entanglement entropy remains a powerful diagnostic tool for characterizing non-Hermitian physics. The entanglement spectrum also reflects the topological characteristics of non-Hermitian topological systems, while unique non-Hermitian entanglement behaviors are also discussed. The review is concluded with several future directions. Through this review,  we hope to provide a useful guide for researchers who are interested in entanglement in non-Hermitian quantum systems.

\end{abstract}
\maketitle


\section{Introduction}
Physics of non-Hermiticity has been observed and analyzed in both classical and quantum systems, including fields such as atomic and nuclear physics, photonics, acoustics, and condensed matter physics~\cite{Moiseyev1998,Daley2014,Miri2019,Weimer2021,Gorini1976,Lindblad1976,Davies1974,Breuer2007}. In quantum systems, non-Hermitian Hamiltonians can be effectively derived from Lindblad quantum master equations which describe the dynamic evolution of open systems interacting with their environment. These Hamiltonians provide a simple and intuitive framework for analyzing complex open systems. One of the most significant achievements in non-Hermitian quantum physics is the study of a class of non-Hermitian systems with parity-time ($\mathcal{PT}$) reversal symmetry possessing stable real spectra~\cite{Bender1998,Bender2007,Bender2018}. This discovery has greatly advanced the study of non-Hermiticity.
In recent years, there has been substantial progress in both the theoretical exploration and experimental investigation of non-Hermitian systems, as demonstrated in the literature, see, e.g., Refs.~\cite{Yuto2020,Bergholtz2021,Zhenbo2015,Xiao2017,Hodaei2017,ShenHuitao2018,Zhu2018,Kunst2018,Wang2019,GoldsteinMoshe2019,Xiao2020,WangXiao2020,LiLinhu2020,LiuYanxia2021,Zhang2021,ZhouZiheng2022,ZhangZhiQiang2022,bian2022,Zhang2022,HanPei2023,YuXueJia2024,CaoKui2023,LiHaowei2023,meden2023symmetric,SunYeyang2024,QianJie2024,ShaoKai2024,ChenJiangzhi2024,ZhongPeigeng2024,XiongYuncheng2024,WangYiTao2024,CaoKui2024} and references therein. Non-Hermitian systems exhibit intriguing physical phenomena, such as the non-Hermitian skin effect~\cite{Yao2018} and exceptional points~\cite{Moiseyev2011,Bergholtz2021,DingKun2022}. Furthermore, the development of non-Bloch band theory has reshaped our understanding of bulk-edge correspondence in 1D non-Hermitian systems, utilizing the concept of a Generalized Brillouin Zone (GBZ)~\cite{Yao2018,Yokomizo2019}.

Entanglement, a concept originating from quantum information theory that quantifies the nonlocal correlations between quantum subsystems, plays a crucial role in condensed matter physics~\cite{Amico2008,Horodecki2009,Nicolas2016}. Over the years, quantum entanglement has become an important perspective for diagnosing fundamental physics and uncovering intriguing emergent phenomena in Hermitian many-body systems, spurring significant progress in the field~\cite{Peschel1999,Chung2000,Chung2001,Cheong20041,Ingo2003,CheongSiew20042,Peschel20041,Alioscia2004,Alexei2005,Levin2006,Li2008,GuZhengCheng2009,Fidkowski2010,Turner2010,Pollmann2010,Hughes2011,Qi2012,Alexandradinata2011,HuangZhoushen20122,Huang2012,Hughes2013,Lee2014,Yangkun2014,Yangkun2014_1,Lee2015free,ZhouYao2023,ZhouYaofractal2023,zhouyaohyperbolic,Mo2023,Lin2024}. 
For example, in gapped systems, entanglement entropy of a quantum many-body ground state obeys the renowned area law~\cite{Hastings2007,Eisert2010}, which is proportional to the area of the subsystem's boundary. In contrast, for $d$-dimensional gapless free-fermion systems, the scaling of entanglement entropy obeys the ``super-area law'', where a logarithmically divergent correction appears in the leading term, and the coefficient of leading term is determined by the geometry of the Fermi surface and the boundary of partition. This scaling law is alternatively called ``Gioev-Klich-Widom scaling'' and can be   physically understood via the Swingle's mode counting arguments~\cite{Widom1990,Gioev2006,Swingle2010}.  Furthermore, the quadratic form of reduced density matrix and translational invariance further provide a convenient way to analytically study entanglement entropy by using    general  properties of Toeplitz matrices, despite the non-sparse nature of correlation matrix~\cite{Ingo2003,Jin2004,Lee2015free}.  As a side,   entanglement on free-fermion systems is studied in $2$D Fermi liquids with interaction~\cite{Yangkun2012}. 
In addition to   fermions,   entanglement entropy in Bose-condensed superfluids or ordered antiferromagnets  includes additive logarithmic corrections associated with the number of Nambu-Goldstone modes~\cite{Metlitski2011}. In topological order, entanglement entropy has a constant term, which is called topological entanglement entropy and encodes the information of the total quantum dimension~\cite{Alioscia2004,Alexei2005,Levin2006} of anyon contents. However,  in some phases without topological orders, e.g., symmetry-protected topological phases with subsystem symmetry,  the extraction of topological entanglement entropy    can suffer from spurious contributions from the nonlocal string order \cite{STEEa,spurious} and this spurious contribution has been extensively studied \cite{STEEb,STEEc,STEEd,zlyprxq24}.  Besides, other entanglement quantities can also capture the intriguing feature of many-body systems. For example, entanglement spectrum, the spectrum of the reduced density matrix, have a connection with the energy spectrum of physical edge states in topological systems~\cite{Li2008,Fidkowski2010,Turner2010,Pollmann2010,Hughes2011,Qi2012,Alexandradinata2011,HuangZhoushen20122,Huang2012}.

Motivated by many exotic phenomena in non-Hermitian quantum systems, using entanglement quantities to study these systems has become an important research topic. Recently, entanglement-related quantities are extensively explored in non-Hermitian systems~\cite{Herviou20192,Herviou2019,Chang2020,Chen2021,Lee2022,Sayyad2021,Dynamics2021,Modak2021,Guo2021,Tu2022,Chen2022Wei,Taberner2022,Dora2022,Chen2022,YiWeiZhu2021,Kawabata2023,Cipolloni2023,Le2023,ChangTse2023,ItableGene2023,Zhou2023,Fossati2023,SticletDoru2023,Wei2023,ZOU2024Deyuan,Orito2023,Zhou20241,Li2024,Qian2024,Zhou2024,Federico2024,Sinha2024,Wen2024,ZhouXin2024,Shunlin2024,MunozArboleda2024,YangZhenghao2024,LiZigeng2024,FengShu2024,LiuJing2024,HuZhang2024,WangXin2024,LiuSirui2024,Soares2024,Korff2007,Bianchini2015,Couvreur2017,LiYaodong2019,LiYue2023,Cheng2024,Tista2023,Jin2024,LuChao2024,Chao2023,Turkeshi2023,Zhou20242,MiaoChuanzheng2024,liu_dynamics}, providing powerful tools for characterizing exotic properties of these systems. 
In this article, we attempt to review recent developments regarding entanglement in non-Hermitian quantum systems. Firstly, in non-Hermitian systems, the definition of the density matrix needs to be modified, involving left and right eigenstates~\cite{Herviou2019,Chang2020}. As a result, the associated entanglement quantities do not necessarily maintain positive definiteness. 
To facilitate comparison, we explore the relationships between entanglement entropy and the correlation matrix in both Hermitian~\cite{Cheong20041,Ingo2003,CheongSiew20042,Peschel20041} and non-Hermitian free-fermion systems~\cite{Herviou2019,Chang2020,Chen2021,Chen2022,Lee2022}. The applications of entanglement quantities in specific models are presented, including using entanglement entropy to describe phases and phase transitions as well as dynamics in non-Hermitian systems~\cite{Chang2020,MunozArboleda2024,YangZhenghao2024,Chen2022,Zhou2023,Zhou20241,Li2024,Wei2023,Shunlin2024,Zhou20242,liu_dynamics}, as well as exotic entanglement phenomena such as negative entanglement entropy~\cite{Chang2020,Lee2022,Wen2024,ZhouXin2024}, the transition of central charge to effective central charge~\cite{Bianchini2015}, complex entanglement spectra~\cite{Chang2020}. 
Due to the complex spectra, biorthogonality, and geometric defectiveness of non-Hermitian matrices, the corresponding entanglement quantities exhibit anomalous behaviors.

The structure of this review is the following. In Sec.~\ref{sec2}, we discuss how to construct non-Hermitian quantum systems, including extracting an effective non-Hermitian Hamiltonian from the Lindblad master equation. In Sec.~\ref{sec3}, we provide a basic knowledge of entanglement in Hermitian systems. Some general properties of entanglement in non-Hermitian systems are  discussed in Sec.~\ref{sec4}, and the novel phenomena in concrete models studies are emphasized in Sec.~\ref{sec5}. In Sec.~\ref{sec6}, we provide a simple conclusion and outlook towards entanglement in non-Hermitian systems. 

\section{Realization of non-Hermitian quantum systems}\label{sec2}

Before discussing quantum entanglement of non-Hermitian systems, we plan to introduce the realization of non-Hermitian effective quantum systems from open quantum systems.
Consider a microscopic system $\mathcal{S}$ coupling to an external environment $\mathcal{E}$, and the full dynamics of the total system is unitary. The general Hamiltonian of the total system is $H_t=H+H_{\mathcal{E}}+H_{I}$ where $H$ is the Hamiltonian of the microscopic system, and $H_{\mathcal{E}}$ is the Hamiltonian of the environment. $H_{I}$ is the interaction Hamiltonian between $\mathcal{S}$ and $\mathcal{E}$, and can be expressed as $H_{I}=g\sum_{\alpha}\hat{\mathcal{S}}_{\alpha}\otimes \hat{\mathcal{E}}_{\alpha} $, where $\hat{\mathcal{S}}$ and $\hat{\mathcal{E}}$ are Hermitian operators and $g$ is the strength of weak system-environment coupling. Here, the total system is described by the total density matrix $\rho_t=|\psi_t\rangle\langle \psi_t|$, where $|\psi_t\rangle$ is the ground state of the total system. The time evolution of the density matrix is characterized by the  {von Neumann equation}, whose partial trace is the dynamic equation of the reduced system, given by $\dot{\rho}=-i\text{Tr}_\mathcal{E}[H_t,\rho_t]$. Here, $\rho=\rho(t)$ is the reduced density matrix by tracing out the environment part of the total density matrix: $\rho=\text{Tr}_\mathcal{E} \rho_t$; $\text{Tr}_\mathcal{E}$ represents tracing out the environment degrees of freedom. After  {Born approximation} that restricts the correlations between system and environment to be small and then $\rho_t\approx \rho(t)\otimes \rho_{\mathcal{E}}$, and  {Markov approximation} that restricts system-environment coupling to be independent of frequency and any correlation functions in the environment to reserve no long-term memory of the coupling with the system, one obtains Markovian quantum master equation. Adding  {rotating-wave approximation} removing terms that oscillate fast with respect to characteristic time scales of the system, one obtains the Lindblad master equation~\cite{Gorini1976,Lindblad1976,Davies1974,Campaioli2024,Wiseman2009}: 
 $\dot{\rho}=-i[H,\rho]+\mathcal{D}(\rho)\equiv \mathcal{L}\rho\,,
 $ where $\mathcal{L}$ is the Liouvillian superoperator. The first term $-i[H,\rho]$ governs the unitary evolution of the system and the second term $\mathcal{D}(\rho)\equiv \mathcal{L}\rho$ is a dissipator. The dissipator is written as
 $ \mathcal{D}(\rho)=\sum_i (\Gamma_i\hat{L}_i\rho\hat{L}_i^{\dag}-\frac{\Gamma_i}{2}\{\hat{L}_i^{\dag}\hat{L}_i,\rho\})\,,
 $ where the first term is quantum jump and the second term describes the coherent non-unitary dissipation of the system. The operators $\{\hat{L}_i\}$ are jump operators; the rate $\{\Gamma_i\}$ is positive. Quantum jumps describe the sudden changes in the state of the system caused by the dissipation and represent the measurement-like action implemented by the environment on the state of the system from the measurement aspect. According to the number of quantum jumps, the stochastic time evolution can be categorized into different quantum trajectories. Taking a stochastic average over trajectories, we may obtain the following equation:
 $\dot{\rho}=-i[(H\!-\!\frac{i}{2}\sum_i\Gamma_i \hat{L}_i^{\dag}\hat{L}_i)\rho -\!\rho (H\!-\!\frac{i}{2}\sum_i\Gamma_i \hat{L}_i^{\dag}\hat{L}_i)^{\dag}]+\sum_i \Gamma_i \hat{L}_i \rho \hat{L}_i^{\dag}\,.
 $ We can define an effective Hamiltonian,
$H_{\text{eff}}=H-\frac{i}{2}\sum_i\Gamma_i \hat{L}_i^{\dag}\hat{L}_i$,
such that the Lindblad master equation can be recast into  $\dot{\rho}=-i[H_{\text{eff}}\rho -\rho H_{\text{eff}}^{\dag}]+\sum_i \Gamma_i \hat{L}_i \rho \hat{L}_i^{\dag}$. Once the quantum jumps can be reasonably ignored, the time evolution equation of $\rho$ formally bears a resemblance to a closed system governed by an effective Hamiltonian  $H_{\text{eff}}$. One way to realize the effective Hamiltonian is to postselect quantum trajectories that do not undergo quantum jumps~\cite{Daley2014}. Hence, the master equation can be considered as averaged over infinitely many trajectories, meaning that all the measurement outcomes are averaged out or discarded. In addition, the master equation also exhibits some phenomena similar to non-Hermitian systems, like Liouvillian superoperator appearing skin effect~\cite{Song2019,Haga2021,YangFan2022,Feng2024} and exceptional points~\cite{Naomichi2019}.

Next, we introduce two prototypical  examples of the realization of non-Hermitian quantum systems from open quantum systems. In the beginning, the simplest open quantum system is a two-level system in which atoms are coupled one by one with an optical cavity field via the electric-dipole interaction~\cite{Imoto1990}. The electric-dipole interaction between the system and measured apparatus is described by the Jaynes-Cumming  interaction, 
$H_{\text{int}}=\gamma(\hat{\sigma} \hat{a}^{\dag}+\hat{\sigma}^{\dag} \hat{a})$, where $\gamma$ is a constant describing the strength of the coupling between the system and measured apparatus, $\hat{a}(\hat{a}^{\dag})$ is the bosonic annihilation (creation) operator, $\hat{\sigma}(\hat{\sigma}^{\dag})$ is the level-lowering (level-raising) operator for the two-level atom defined by $\hat{\sigma}\equiv|g\rangle\langle e|(\hat{\sigma}^{\dag}\equiv|e\rangle\langle g|)$, where $|g\rangle$ and $|e\rangle$ are state vectors for the ground and excited states, respectively. Here, we suppose this two-level system at the small photon-number regime, where at most a single photon is absorbed by the detector. Then, the spontaneous emission of a photon is described by a single jump operator, $\hat{L}= \hat{\sigma}$. When the system is in the excited states, due to the spontaneous emission, we should detect a photon. If the system has no photons detected by the  apparatus, which means the absence of a jump process, the system should be in the ground states~\cite{Imoto1990,Yuto2020}.  Correspondingly, the time evolution can be described by the effective non-Hermitian Hamiltonian:
 $ H_{\text{eff}}=H-\frac{i\Gamma}{2} \hat{\sigma}^{\dag} \hat{\sigma}, 
 $  where $H$ describes the internal dynamics of the two-level system, $\Gamma$ characterizes the decay rate, and $\hat{\sigma}^{\dag} \hat{\sigma}$ presents a projection onto the excited state. When the jump process is probed,  the system is projected onto the ground state after experiencing spontaneous emission of a photon. When averaging out or discarding all measurement outcomes, the time evolution becomes the master equation~\cite{Daley2014}.

The second example is related to a $1$D free-fermion system, given by $H=-\sum_i \frac{t}{4}(\hat{c}_{i+1}^{\dag}\hat{c}_i+\hat{c}_i^{\dag}\hat{c}_{i+1})$, where $\hat{c}^{\dag}_i(\hat{c}_i)$ is the creation (annihilation) operator at site $i$. The system goes through a projective measurement and the observable is the occupation number of local quasimodes about right-moving wave packet. The measurement is realized by two-site projectors, $P_i=\hat{\xi}_i^{\dag}\hat{\xi}_i$ in which $\hat{\xi}_i^{\dag}=\frac{1}{\sqrt{2}}(\hat{c}_i^{\dag}-i\hat{c}_{i+1}^{\dag})$ can be regarded as the creation operator of a right-moving wave packet. Then, the jump operator is $\hat{L}_i=P_i$. If one neglects the quantum jumps, the master equation can be described by the effective non-Hermitian Hamiltonian~\cite{FengXu2023, WangYuPeng2024, Feng2024}, given by
 $H_{\text{eff}}=\frac{1}{4}\sum_i[(-t+\Gamma)\hat{c}_i^{\dag}\hat{c}_{i+1}-(t+\Gamma)\hat{c}_{i+1}^{\dag}\hat{c}_i-i\Gamma(n_i+n_{i+1})]\,.
 $ This $H_{\text{eff}}$ can be regarded as a special case of the Hatano-Nelson model~\cite{Hatano1996} and exhibits non-Hermitian skin effect under open boundary conditions, which induces abundant dynamical phenomena, like unconventional reflection and entanglement suppression~\cite{Haoshu2022,Kawabata2023}.

Remarkably, significant progress has been made in experimental realization and measurement. For example, non-Hermitian bulk-boundary correspondence is verified by directly measuring topological edge states and the skin effect in discrete-time non-unitary quantum-walk dynamics of single photons~\cite{Xiao2020}, while higher-order skin effects are observed in coupled resonator acoustic waveguides~\cite{Zhang2021}. Exceptional points are also simulated in the experiment. Researchers report the observation of exceptional points in a photonic crystal slab~\cite{Zhenbo2015}, and higher-order exceptional points in a coupled cavity arrangement with a precisely engineered gain-loss distribution~\cite{Hodaei2017}. While there are many other intriguing experimental results, space limitations prevent a comprehensive discussion here and we will mainly focus on the entanglement perspective of non-Hermitian free-fermion systems.

\section{A short review on entanglement in Hermitian systems}\label{sec3}
\begin{figure}[htbp]
\centering
\includegraphics[width=8.5cm]{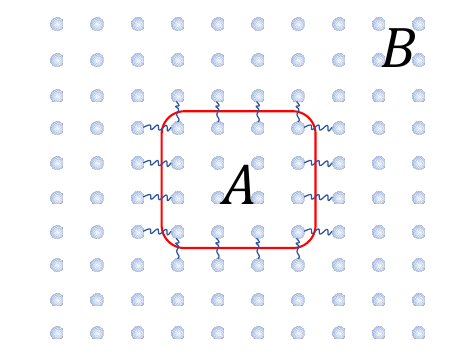}
\caption{A physical system is divided into two subsystems $A$ and $B$ for the purpose of quantifying the entanglement between the two subsystems. The red curve denote the entanglement cut that separate the two subsystems. The wavy lines denote the quantum correlation induced by local hopping or local interactions.}
\label{bipartition}
\end{figure}

To measure quantum correlation from quantum information perspective, entanglement entropy  is a useful and suitable quantity in various quantum systems. Consider a quantum lattice system with a density matrix denoted by $\rho$, defined as $\rho = |GS\rangle\langle GS|$, where $|GS\rangle$ represents the quantum many-body ground state. The system is partitioned into two subsystems, namely, $A$ and $B$ as illustrated in Fig.~\ref{bipartition}. By tracing out subsystem $B$,  we can end up with the reduced density matrix  $\rho_A = \text{Tr}_B \rho$. The $n$th-order R\'enyi entropy is then defined as $S_n = \frac{1}{1-n} \ln \text{Tr} (\rho_A^n)$. As $n$ approaches 1, this expression converges to the von Neumann entropy, often referred to as entanglement entropy~\cite{Bennett1996,Pasquale2004}, which is defined as $S = -\text{Tr}(\rho_A \ln \rho_A)$.

Despite being non-interacting, free-fermion systems possess rich entanglement phenomena due to the presence of  Fermi statistics. Thanks to the non-interacting nature, there is a  convenient   way to study entanglement entropy and entanglement spectrum through  the single-particle correlation matrix~\cite{Cheong20041,Ingo2003,CheongSiew20042,Peschel20041}. In the supplementary note of Ref.~\cite{Lin2024}, a  concrete derivation about entanglement entropy and correlation matrix in Hermitian free-fermion systems is provided. For free-fermion systems, the second-quantized Hamiltonian is of quadratic form: $H=\sum_{ij}\hat{c}_i^{\dag}\mathcal{H}_{i,j}\hat{c}_j$, where $i,j,\cdots$ label lattice sites as well as other indices (e.g., spin, orbitals) at each site. $\hat{c}_i^{\dag}$ and $\hat{c}_j$ are respectively  fermionic creation and annihilation operators which satisfy the standard anticommutation relations: $\{\hat{c}_i,\hat{c}_j^{\dag}\}=\delta_{i,j}$ and $\{\hat{c}_i,\hat{c}_j\}=\{\hat{c}_i^{\dag},\hat{c}_j^{\dag}\}=0$. The kernel matrix $\mathcal{H}$ can be diagonalized by a set of ortho-normal eigenstates $\{|\alpha\rangle\}$ satisfying $\langle\alpha|\beta\rangle=\delta_{\alpha,\beta}$, and becomes $\mathcal{H}=\sum_{\alpha}E_{\alpha}|\alpha\rangle\langle\alpha|$. Then, the  Hamiltonian can be written in a diagonal form
 $H=  \sum_{\alpha}E_{\alpha}\hat{\psi}^{\dag}_{\alpha}\hat{\psi}_{\alpha}\,,
  $ where $\hat{\psi}^{\dag}_{\alpha}$ and $\hat{\psi}_{\alpha})$ are also  fermionic: $|\alpha\rangle=\hat{\psi}^{\dag}_{\alpha}|0\rangle$, $\hat{\psi}_{\alpha}|0\rangle=0$ and $\{\hat{\psi}_{\alpha},\hat{\psi}_{\beta}^{\dag}\}=\delta_{\alpha,\beta}$. $|0\rangle$ is the vacuum state without fermions. In addition, two creation operators can be   transformed to each other via
 $ \hat{\psi}^{\dag}_{\alpha}=\sum_i \psi_{\alpha}(i)\hat{c}_i^{\dag}$ and $
\hat{c}_i^{\dag}=\sum_{\alpha}\psi^*_{\alpha}(i)\hat{\psi}^{\dag}_{\alpha}\,,
  $ where the wavefunction $\psi_{\alpha}(i)$ is  defined as $\psi_{\alpha}(i)=\langle i|\alpha\rangle$. The many-body ground state of the original free-fermion system with $N$ fermions can be constructed as: $|GS\rangle=\prod_{\alpha\in occ.}\hat{\psi}^{\dag}_{\alpha}|0\rangle$, where $\alpha\in occ.$ means that  $N$ single-particle states with lowest energies are occupied. Correspondingly, the density matrix is expressed as $\rho=|GS\rangle\langle GS|$ and the reduced density matrix is $\rho_A=\text{Tr}_{B}\rho$. As for the reduced density matrix of free-fermion systems, due to the validity of Wick contraction in both subsystem and original system,  it also has the quadratic form~\cite{Peschel1999,Chung2000,Chung2001}:
 $ \rho_A= \mathcal{Z}^{-1}e^{-H^E}$ with the \textit{entanglement ``Hamiltonian''} $H^E=\sum_{i,j\in A}\hat{c}_i^{\dag}h_{i,j}^E\hat{c}_j$. Here,  $\mathcal{Z}=\text{Tr}(e^{-H^E})$ is a normalization constant and $h^E$ is the kernel matrix of $H^E$. Similar to the system's Hamiltonian $H$, we may introduce $\{\phi_n(i)\}$ as the eigenfunctions of  $h^E$ with eigenvalues $\{\xi_n\}$, such that $h^E|\phi_n\rangle=\xi_n$ and $\langle i|\phi_n\rangle=\phi_n(i)$. Then, the fermionic operator $\hat{c}_i$ can be transformed to a new set of fermionic operators $\hat{a}_n$ via
 $ \hat{c}_i=\sum_n \phi_n(i)\hat{a}_n\,.
 $ Under these transformations with $\sum_{i}\phi_n(i)\phi_{n'}^*(i)=\delta_{n,n'}$ and $\sum_n \phi_n^*(i)\phi_n(j)=\delta_{i,j}$, we can diagonalize the kernel matrix $h_{i,j}^E=\sum_n \phi_n(i)\phi_n^*(j)\xi_n$. Meanwhile, $\rho_A$   can be sent  into a diagonal form
 $  
\rho_A=\mathcal{Z}^{-1}e^{-\sum_n\xi_n \hat{a}_n^{\dag}\hat{a}_n}\,,
 $ where the set of real numbers $\{\xi_n\}$ forms the  {single-particle entanglement spectrum}. The wavefunctions $\{\phi_n(i)\}$ are called ``entanglement wavefunctions'' or ``Schmidt vectors'', which are applied to a variety of many-body systems~\cite{Peschel2009,vidal2020,ZhouYao2023}.

Interestingly, there is an exact relation   between the correlation matrix and the entanglement Hamiltonian matrix~\cite{Ingo2003}. For a free-fermion system, the definition of a real-space correlation matrix is
 $ C^A_{i,j}=\langle GS|\hat{c}_j^{\dag}\hat{c}_i|GS\rangle\,,
 $ where   $i,j,\cdots$ are restricted in subsystem $A$. 
 In addition, the correlation matrix can be written by density matrix as
 $ C^A_{i,j}=\text{Tr}_A(\rho_A \hat{c}_j^{\dag}\hat{c}_i)=\text{Tr}_A[\text{Tr}_B(\rho\hat{c}_j^{\dag}\hat{c}_i)]\,.
 $ The correlation matrix restricted in the subsystem $A$ can be diagonalized as
 $ C^A_{i,j}
 =\sum_n \phi_n^*(j)\phi_n(i)   ({e^{\xi_n}+1})^{-1}\,.
 $ Correspondingly, the eigenvalue, denoted as $\varepsilon_n$, of  $C_{i,j}^A$ is fully determined by the entanglement spectrum via the identity 
 $ \varepsilon_n=  ({e^{\xi_n}+1})^{-1}
 $.  Formally, the exact relation between the two matrices can be expressed in a compact form, $h^E=\ln[(C^A)^{-1}-\mathbb{I}]$, where $\mathbb{I}$ is the identity matrix. Due to the above one-to-one correspondence between  $\varepsilon_n$ and $\xi_n$,  $\varepsilon_n$ is more often referred to as the single-particle entanglement spectrum, which has been a well-accepted convention in the literature of free-fermion entanglement. Since $\varepsilon_n$ is restricted within the range of [0,1], this convention brings   convenience to both analytic and numerical calculations. By means of $\{\varepsilon_n\}$, the entanglement entropy of a free-fermion system can be expressed by the eigenvalues of the correlation matrix,
$ S
=-\sum_n[\varepsilon_n\ln\varepsilon_n+(1-\varepsilon_n)\ln(1-\varepsilon_n)]\,.
 $ Obviously, $\varepsilon_n=0.5$ is very special as it provides the maximal entanglement contribution and corresponds to zero modes of entanglement Hamiltonian. In contrast, the eigen modes with either $\varepsilon_n=0$ or  $\varepsilon_n=1$  do not contribute to entanglement entropy.

Another property of the correlation matrix  $C^A$ is that  it can be rigorously determined  by two mutually non-commuting projectors~\cite{Alexandradinata2011, HuangZhoushen20122,Huang2012, Lee2014, Lee2015free}, namely, $C^A=\mathcal{RPR}$. Here,  the ``real-space projector'' $\mathcal{R}=\sum_{i\in A}|i\rangle\langle i|$. The ``momentum-space projector'' $\mathcal{P}$ projects out all unoccupied states through $\mathcal{P}=\sum_{\alpha}\theta(-\varepsilon_\alpha)|\alpha\rangle\langle \alpha|$, where $|\alpha\rangle$ and $\varepsilon_\alpha$ respectively represent the eigenstates and eigenvalues of  the kernal matrix $\mathcal{H}_{i,j}$ in the free-fermion Hamiltonian $H$; the symbol $\theta$ denotes  the standard step function.  Take two typical examples, $\mathcal{P}=\sum_k\theta(-\varepsilon_k)$ for a fermi sea of fermi gas, and $\mathcal{P}=\frac{1}{2}[\mathbb{I}-\hat{d}(k)\cdot \sigma]$ for a two-band free-fermion model with Hamiltonian $H=d(k)\cdot \sigma$ where $ \sigma$ are the Pauli matrices.  Based on Spec$(\mathcal{R}\mathcal{P} \mathcal{R})$ = Spec$(\mathcal{P}\mathcal{R}\mathcal{P})$\footnote{``Spec($\cdots$)'' means the spectrum of the matrix.}, a rigorous position-momentum duality can be established in which the   entanglement spectrum  keeps invariant~\cite{ Lee2014}. This duality can provide a physical understanding for the scaling of entanglement entropy of the typical excited state in free-fermion systems~\cite{Yangkun2015}.

Apart from entanglement entropy, the entanglement spectrum provides a new angle to characterize the ground state property.  As one of the important developments, the Ref.~\cite{Li2008}   finds the correspondence between the entanglement spectrum and edge mode spectrum of the fractional quantum Hall states. Entanglement also plays a crucial role in characterizing symmetry-protected topological phases \cite{GuZhengCheng2009,Pollmann2010}. In topological free-fermion systems, e.g.,  topological insulators, topological superconductors,  and higher-order Weyl semimetals, entanglement spectrum exhibiting 1/2 modes~\cite{Fidkowski2010,ZhouYao2023} displays a quantum informative signature of the edge states. Besides, the degeneracy of the entanglement spectrum can reflect the non-trivial topological nature, such as the double degeneracy of the entanglement spectrum of the Haldane phase~\cite{GuZhengCheng2009,Pollmann2010}. Towards the understanding on the exact relation between edge modes and entanglement spectrum in topological systems, several theoretical attempts have been made~\cite{Hughes2011,Qi2012}.

\section{General properties of entanglement in non-Hermitian systems}\label{sec4}

In the early stages, researchers discussed the entanglement entropy of non-unitary conformal field theory (CFT) and generalized the entanglement entropy to non-Hermitian systems by introducing the left and right many-body ground states (denoted as $|G_L\rangle$ and $|G_R\rangle$) to define the density matrix~\cite{Bianchini2015,Couvreur2017}. Later, in non-Hermitian free-fermion systems, the density matrix is widely defined as $\rho=|G_R\rangle\langle G_L|$ with $\rho^{\dag}\neq \rho$, and entanglement entropy still maintains the original definition. Entanglement entropy can also characterize phases and phase transitions of a part of non-Hermitian systems. However, the eigenvalues of the reduced density matrix are no longer exclusively positive semi-definite, so entanglement entropy may  become negative or even complex~\cite{Herviou2019,Chang2020,Lee2022,Wen2024}, which leads to intricate interpretation in terms of quantum information. 
Faced with these issues, some researchers try to modify the definition of reduced density matrix and entanglement entropy to find the most  appropriate way to characterize non-Hermitian physics.

The first approach uses the original definition, directly extended from the Hermitian case, 
\begin{equation}
S=-\text{Tr} \rho_A \ln \rho_A
\end{equation}
where $\rho_A=\text{Tr}_B \rho$  with an unusual form of density matrix $\rho=|G_R\rangle\langle G_L|$~\cite{Herviou20192,Herviou2019, Chang2020, Chen2021, Lee2022, Modak2021, Guo2021, Taberner2022,Chen2022,Zhou2023,Chen2022Wei, Wei2023,Wen2024,Shunlin2024,MunozArboleda2024,YangZhenghao2024}. Here, $|G_R\rangle$ and $|G_L\rangle$  are biorthogonal ground states to be defined below shortly~\cite{Brody2013}. Compared to other definitions $\rho=|G_R\rangle\langle G_R|$ or $\rho=|G_L\rangle\langle G_L|$, it can more comprehensively reflect the intrinsic property of static non-Hermitian systems. For the entanglement Hamiltonian of biorthogonal density matrix, different symmetries of original Hamiltonian  can be realized~\cite{Herviou2019}. In particular, this definition is sensitive to criticality and provides information about the non-unitary CFT~\cite{Chang2020,Lee2022,Wen2024}. Besides, the entanglement spectrum characterizes the topological properties of the initial Hamiltonian, provided that the system possesses separable bands~\cite{Herviou2019} .

Concretely,   for a generic quadratic non-Hermitian Hamiltonian, $H=\sum_{ij}\hat{c}_i^{\dag} \mathcal{H}_{ij}\hat{c}_j$ with a non-Hermitian kernel matrix $\mathcal{H}\neq \mathcal{H}^{\dag}$.  Fermionic operators still satisfy $\{\hat{c}_i,\hat{c}_j^{\dag}\}=\delta_{ij}$ and $\{\hat{c}_i,\hat{c}_j\}=\{\hat{c}_i^{\dag},\hat{c}_j^{\dag}\}=0$. The kernel matrix $\mathcal{H}$ can be diagonalized by a set of biorthogonal eigenstates $\{|R_{\alpha}\rangle, |L_{\alpha}\rangle\}$, and they satisfy biorthogonal relation $\langle L_{\alpha}|R_{\beta}\rangle=\delta_{\alpha\beta}$, where  $\mathcal{H}|R_{\alpha}\rangle=\epsilon_{\alpha}|R_{\alpha}\rangle$, $\mathcal{H}^{\dag}|L_{\alpha}\rangle=\epsilon_{\alpha}^*|L_{\alpha}\rangle$. Here, $\{|R_{\alpha}\rangle\}$  and $\{|L_{\alpha}\rangle\}$ are respectively right and left eigenvalues.  The right (left) eigenstates are not mutually orthogonal, and the strength of non-orthogonality between the right (left) eigenstates can be measured by the Petermann factor and related variants \cite{Petermann1979, WangHeming2020,Zou2023}.  By means of biorthogonal eigenbases, the kernel matrix $\mathcal{H}$ can be diagonalized: $\mathcal{H}=\sum_{\alpha}\epsilon_{\alpha}|R_{\alpha}\rangle\langle L_{\alpha}|$ such that the  quadratic non-Hermitian  Hamiltonian $H$ is sent to a diagonalized form: 
 $  H=\!\!\sum_{\alpha}\epsilon_{\alpha}\!\!\sum_i (R_{\alpha i}\hat{c}_i^{\dag}) \sum_j(L^*_{\alpha j}\hat{c}_j)=\!\!\sum_{\alpha}\epsilon_{\alpha} \hat{\psi}_{R\alpha}^{\dag} \hat{\psi}_{L\alpha}\,,
 $ 
where $\hat{\psi}_{R\alpha}^{\dag}$ and $\hat{\psi}_{L\alpha}^{\dag}$ are respectively the right and left fermionic creation operators satisfying $|R_{\alpha}\rangle=\hat{\psi}_{R\alpha}^{\dag}|0\rangle$, $|L_{\alpha}\rangle=\hat{\psi}_{L\alpha}^{\dag}|0\rangle$, and $\hat{\psi}_{R\alpha}|0\rangle=\hat{\psi}_{L\alpha}|0\rangle=0$. Especially, they satisfy  $\{\hat{\psi}_{L\alpha},\hat{\psi}_{R\beta}^{\dag}\}=\delta_{\alpha\beta}$, $\hat{\psi}_{L\alpha}$ and $\hat{\psi}_{R\alpha}$ are called the bi-fermionic operators. In addition, two kinds of creation operators can be   transformed into each other via
 $ \hat{\psi}_{R\leftrightarrow L,\alpha}^{\dag}=\sum_i \langle i|(R\leftrightarrow L)_{\alpha}\rangle \hat{c}_i^{\dag}=\sum_i (R\leftrightarrow L)_{\alpha i}\hat{c}_i^{\dag}$, 
 $\hat{c}_i^{\dag}=\sum_{\alpha}\langle L_{\alpha}|i\rangle\hat{\psi}_{R\alpha}^{\dag}=\sum_i L^*_{\alpha i}\hat{\psi}_{R\alpha}^{\dag}\,$, 
and $\hat{c}_j=\sum_{\alpha}\langle j|R_{\alpha}\rangle\hat{\psi}_{L\alpha}=\sum_j R_{\alpha j}\hat{\psi}_{L\alpha}$, where $R_{\alpha i}$  is   defined as $R_{\alpha i}=\langle i|R_{\alpha}\rangle$.

Due to the biorthogonal eigenstates and the complex energy spectra, the definitions of many-body ground states of a non-Hermitian free-fermion system with $N$ fermions become various. A many-body right ground state can be constructed as $|G_R\rangle=\prod_{\alpha\in \text{occ.}}\hat{\psi}_{R\alpha}^{\dag}|0\rangle$ satisfying $H|G_R\rangle=\sum_{\alpha\in \text{occ.}} \epsilon_{\alpha}|G_{R}\rangle$. Similarly, a many-body left ground state is $|G_L\rangle=\prod_{\alpha\in \text{occ.}}\hat{\psi}_{L\alpha}^{\dag}|0\rangle$ satisfying $H^{\dag}|G_L\rangle=\sum_{\alpha\in \text{occ.}} \epsilon^*_{\alpha}|G_{L}\rangle$. Due to the complex energy spectra of non-Hermitian systems, the occupied energy ``occ'' also has many selections, like real spectra, imaginary spectra, or modulus of spectra, and many  researchers adopt states in which occupation is defined through  ordering the real part of energies from bottom. Naturally, a many-body density matrix defined by right and left ground state can be written as:
 $\rho=|G_R\rangle\langle G_L|
 $ which leads to $\rho^{\dag}\neq \rho$ and $\rho^2=\rho$ via the properties of biorthogonality. 

By   partitioning the whole system into two subsystems $A$ and $B$, and subsequently tracing out subsystem $B$, we can deduce the reduced density matrix: $\rho_A=\text{Tr}_B \rho$ by definition. Similar to Hermitian cases, here $\rho_A$ can also be written in quadratic form: 
 $ \rho_A=\mathcal{Z}^{-1}e^{-H^E}, H^E=\sum_{i,j}\hat{c}_i^{\dag} h_{ij}^E \hat{c}_j\,,
 $ where entanglement Hamiltonian matrix $h^E$ is non-Hermitian, $\mathcal{Z}=\text{Tr}(e^{-H^E})$ is a normalization constant. Let $|\varphi_{Rn}\rangle$ be the right eigenvector of the entanglement Hamiltonian matrix $h^E$ with eigenvalue $\xi_n$, $|\varphi_{Ln}\rangle$ be the left eigenvector with eigenvalue $\xi_n^*$, such that  $h^E|\varphi_{Rn} \rangle=\xi_n|\varphi_{Rn} \rangle$, $(h^E)^{\dag}|\varphi_{Ln} \rangle=\xi^*_n|\varphi_{Ln} \rangle$, $\langle i|\varphi_{Rn}\rangle=\varphi_{Rn}(i)$ and $\langle j|\varphi_{Ln}\rangle=\varphi_{Ln}(j)$. As a result, $\rho_A$ can be written into diagonal form
 $  
\rho_A=\frac{1}{Z}e^{-\sum_n\xi_n \hat{\psi}_{Rn}^{\dag}\hat{\psi}_{Ln}}\,,
 $ where $\hat{\psi}_{Rn}^{\dag}=\sum_i \varphi_{Rn}(i)\hat{c}_{i}^{\dag}$, $\hat{\psi}_{Ln}=\sum_j \varphi^*_{Ln}(j)\hat{c}_{j}$ , and $h^E_{i,j}=\sum_n \xi_n\varphi_{Rn}(i) \varphi^*_{Ln}(j)$,  based on the relationship $\hat{c}_i^{\dag}=\sum_n \varphi^*_{Ln}(i)\hat{\psi}_{Rn}^{\dag},\hat{c}_i=\sum_n \varphi_{Rn}(i)\hat{\psi}_{Ln}$, $\sum_{i}\varphi_{Rn}(i)\varphi^*_{Ln'}(i)=\delta_{n,n'}$,  and $\sum_{n}\varphi_{Rn}(i)\varphi^*_{Ln}(j)=\delta_{ij}$.

By means of biorthogonal ground states,    the real-space correlation matrix now  is defined as 
 $ C^A_{i,j}=\langle G_L|\hat{c}_j^{\dag}\hat{c}_i|G_R\rangle=\text{Tr} \rho_A\hat{c}_j^{\dag}\hat{c}_i\,,
 $ where $i(j)$ is restricted in   subsystem $A$. Due to the biorthogonality and anti-commutation relation of left and right fermionic creation and annihilation operators, the fermions still satisfy Fermi statistics in non-Hermitian free-fermion systems~\cite{Herviou2019}.
  Then, the correlation matrix restricted in subsystem $A$ can be diagonalized by the eigenbasis of the entanglement Hamiltonian,
$C^A_{i,j}\!\! = \text{Tr}(\rho_A \hat{c}_j^{\dag}\hat{c}_i)
 =\!\sum_n\varphi^*_{Ln}(j) \varphi_{Rn}(i)  ({e^{\xi_n}+1})^{-1}\,.
 $   The eigenvalues $\varepsilon_n$ of the correlation matrix are fully determined by the entanglement spectrum via this identity,
 $ \varepsilon_n= ({e^{\xi_n}+1})^{-1}\,.
 $ Correspondingly, in non-Hermitian systems, there also exists an exact relation between two matrices, $h^E=\ln[(C^A)^{-1}-\mathbb{I}]$.  Entanglement entropy can be expressed by the eigenvalue $\varepsilon_n$ of the correlation matrix: 
 \begin{equation}
 S=-\sum_n[\varepsilon_n\ln \varepsilon_n+(1-\varepsilon_n)\ln(1-\varepsilon_n)]\,.
 \end{equation}

The above discussion on entanglement entropy is still based on the standard definition, i.e., $S_A=-\text{Tr}\rho_A \ln \rho_A$. Below we introduce two   modified definitions that have appeared in the literature. The first modification involves a redefinition of the reduced density matrix and entanglement entropy using a modified trace, which is model-dependent. The form of this modified trace is determined by both geometrical and quantum group considerations~\cite{Korff2007, Couvreur2017}. For example, in the study of the critical non-Hermitian XXZ spin chain, the modified entanglement entropy is
\begin{equation}
S = -\text{Tr} (q^{2\sigma_A^z} \rho_A \ln \rho_A),
\end{equation}
where $\rho_A = \Tr_B(q^{-2\sigma_B^z} \rho)$, with $\rho \equiv |0_R\rangle \langle 0_L|$, where $q$ is a coupling coefficient factor in the Hamiltonian, and $\sigma_B^z$ is the Pauli matrix acting on the sites in subsystem $B$. Here, $|0_R\rangle$ ($|0_L\rangle$) represents the right (left) many-body ground state.  Researchers calculate the entanglement entropy for different definitions.  The entanglement entropy for the modified version is $S = \ln(q + q^{-1})$, which incorporates the coupling coefficient factor $q$ from the non-Hermitian term. For the standard entanglement entropy, $S = \ln 2$ when $\rho = |0_R\rangle \langle 0_R|$, and $S = -\Tr (\rho_A \ln \rho_A)$, the result matches the entanglement entropy of the critical Hermitian XXZ spin chain (with $q = 1$), showing no signature of non-Hermiticity. The density matrix defined by only right eigenstates reflects properties of the many-body right eigenstates themselves~\cite{Herviou2019} . However, this definition is extensively studied in the entanglement dynamics of non-Hermitian dissipative systems without quantum jumps~\cite{Dynamics2021,Kawabata2023,Le2023,Orito2023,Li2024,Zhou2024,LiZigeng2024,Soares2024}.

The second type of modified entanglement entropy is model-independent and is defined as 

\begin{equation}
S = -\text{Tr} (\rho_A \ln |\rho_A|)  
\end{equation}
and the $n$-th modified R\'enyi entropy is $S^{(n)} = \frac{1}{1-n} \ln(\text{Tr}(\rho_A |\rho_A|^{n-1}))$, as proposed in Ref.~\cite{Tu2022}. This approach is based on the expected value of the measure. The authors demonstrate that these entanglement quantities, which they consider as generic entanglement and R\'enyi entropies for both Hermitian and non-Hermitian critical systems, yield numerical results of negative central charges consistent with predictions from non-unitary CFT. They also clarify that this type of definition is equivalent to the first type, which employs the modified trace formalism, in quantum group symmetric spin models. In Ref.~\cite{Cipolloni2023}, this type of entanglement entropy is applied to describe non-Hermitian many-body quantum chaos, modeled by the Ginibre ensemble, where they observe significant suppression in the entanglement entropy of typical eigenstates. Ref.~\cite{Fossati2023} utilizes this modified entanglement entropy to characterize symmetry-resolved entanglement in the ground state of the non-Hermitian Su-Schrieffer-Heeger (SSH) chain at the critical point. The scaling limit of this system corresponds to a $bc$-ghost non-unitary CFT.  This modified entanglement entropy is also studied in the non-Hermitian SSH model from the perspective of  GBZ~\cite{YangZhenghao2024}.

Similar to Hermitian systems, the correlation matrix can be expressed by projectors, $C^A=\mathcal{RPR}$. Due to $\mathcal{P}^{2}= \mathcal{P},\mathcal{R}^{2}= \mathcal{R}$, one can prove the invariance of spectrum: Spec$(\mathcal{R}\mathcal{P} \mathcal{R})$ = Spec$(\mathcal{P}\mathcal{R}\mathcal{P})$ in non-Hermitian systems.  Correspondingly, a position-momentum duality is studied in non-Hermitian systems, which have a complete set of biorthonormal eigenvectors and an entirely real energy spectrum \cite{Chen2021}. Position-momentum duality preserves the entanglement spectrum, as indicated by the equality: 
\begin{equation}
\text{Spec}(\mathcal{R}_o\mathcal{P}_o \mathcal{R}_o) = \text{Spec}(\mathcal{R}_d\mathcal{P}_d\mathcal{R}_d)\,,
\end{equation}
 where $\mathcal{R}$ denotes the real-space projector restricted to subsystem $A$ and $\mathcal{P}$ signifies the Fock-space projector restricted to occupied states. The subscript $o$ denotes the original system, while the subscript $d$ denotes the dual system. They delineate two types of non-Hermitian models based on system properties.

The rigorous duality between a non-Hermitian non-interacting Hamiltonian $H_o$ and its dual Hamiltonian $H_d$ involves two key steps, as shown in Fig.~1 in Ref.~\cite{Chen2021}. In the first step, to physically ensure that the real-space projectors are Hermitian before and after duality, a similarity transformation $\mathcal{O}$ is applied to both the Fock-space projector $\mathcal{P}_o$ and the real-space projector $\mathcal{R}_o$. In the second step, the real space and Fock space are interchanged along with Fermi surface and partition exchanging, 
naturally defining two new projectors $\mathcal{R}_d$ and $\mathcal{P}_d$ for the dual system. This duality preserves the entanglement spectrum, i.e., Spec$(h_o^E)$=Spec$(h_d^E)$, where the entanglement Hamiltonian is defined as $h^E=\ln[(RPR)^{-1}-\mathbb{I}]$. This mapping transforms a non-Hermitian system $H_o$ into a new one $H_d$ while maintaining identical entanglement spectrum and entanglement entropy. Therefore, the entanglement properties of $H_o$ can be analyzed by studying $H_d$.  If the dual system $H_d$ is found to be Hermitian, it implies that non-Hermiticity does not play a significant role in the entanglement of $H_o$. The condition for $H_d$ to be Hermitian is the existence of such a similarity transformation operator $\mathcal{O}$ satisfying $\mathcal{O}^{-1}R_o \mathcal{O}=\mathcal{O}^{\dag}\mathcal{R}_o \mathcal{O}^{\dag -1}$. Thus, if at least such a similarity transformation exists for a given $H_o$, the system is classified as type I. Otherwise, it is categorized as type II.

 An example of type I is the nonreciprocal model,
 \begin{equation}
 H_o=-t\sum_{x=1}^{L} (e^{\alpha}\hat{c}_x^{\dag} \hat{c}_{x+1}+e^{-\alpha}\hat{c}_{x+1}^{\dag}\hat{c}_{x})\,.
 \end{equation}
   The nonreciprocal left/right hopping $te^{\pm \alpha}$ can arise from asymmetric gain/loss. Under open boundary conditions,  the right and left eigenstates can be exactly written down. Choosing half-filling of the system and a partition is a half-chain. Applying two steps of duality to the $H_o$, we obtain
\begin{align}
H_d=-t'\sum_{x=1}^{L} (\hat{c}_x^{\dag}\hat{c}_{x+1}+\hat{c}_{x+1}^{\dag}\hat{c}_{x})\,,
 \end{align} where $t'$ is the hopping integral and its strength doesn't influence the entanglement. The dual Hermitian Hamiltonian does not depend on the parameter $\alpha$, which indicates that non-Hermiticity plays no role in entanglement entropy and entanglement spectrum  in the original non-Hermitian system. The partition of the dual system is a half-chain. As for a general partition of the original system, the Fermi surface of the dual system can be tuned by introducing a chemical potential.

An example of type II is the non-Hermitian SSH model on a bipartite lattice at half-filling,
 \begin{equation}
 \begin{split}
 H_o=&\sum_{x=1}^{N-1}(\omega \hat{c}_{2x}^{\dag}\hat{c}_{2x+1}+\upsilon \hat{c}_{2x+1}^{\dag}\hat{c}_{2x+2})+h.c.\nonumber \\
&+\sum_{x=1}^Niu(\hat{c}_{2x}^{\dag}\hat{c}_{2x}-\hat{c}_{2x+1}^{\dag}\hat{c}_{2x+1})\,
\end{split}
\end{equation} 
   with $u,\upsilon,\omega\in \mathbb{R}$. Under periodic boundary conditions, the system is $\mathcal{PT}$-symmetric and they restrict their study to the region of the real spectrum. Then, perform two steps of the duality to the original system. The dual Hamiltonian has a general form 
   \begin{equation}
   H_d=\sum_{k,\ell}\epsilon_{k,\ell}\psi^{\dag}_{r,k,\ell}\psi_{l,k,\ell}\,,
\end{equation}    where $\epsilon_{k,\ell}$ is the dispersion relation with $\epsilon_{k,\ell}<0$ for $k\in \mathcal{A}_o,\ell=\pm$ can be interpreted as internal degrees or layer indices, $\mathcal{A}_o$ is the partition of the original system and $ \psi^{\dag}_{r,k,\ell}(\psi_{l,k,\ell})$  are bifermionic operators. When $\mathcal{A}_o$ is half the chain and the dispersion relation should satisfy $\epsilon_{k,
\ell}=-2 t \sqrt{N}\cos k a$, then the dual Hamiltonian can be written as
 $H_d=-t\sum_{x}\sum_{y=x\pm a} \hat{c}_{x}^{\dag} e^{i \mathbf{A}_{x,y}\cdot \sigma+iA_{x,y}^0 \sigma_0}\hat{c}_y\,,
 $ where $\hat{c}_x=(\hat{c}_{x,-},\hat{c}_{x,+})^T$ is a two component spinor, $\sigma=(\sigma_x,\sigma_y,\sigma_z)$ is a vector of Pauli matrices and $\sigma_0$ is an identity matrix. The fields $\mathbf{A}_{x,y}$ and $A_{x,y}^0$ reside at the link $(x,y)$ and no longer keep anti-symmetric on its spatial indices, $\mathbf{A}_{x,y}\neq - \mathbf{A}_{y,x}, A_{x,y}^0\neq - A_{y,x}^0$. They map the non-Hermitian SSH model to non-Hermitian non-Abelian gauge field theory.

\section{Concrete studies}\label{sec5}
We have discussed the general properties of entanglement in non-Hermitian systems above. In this section, we introduce various applications in   concrete models. Examples include the non-Hermitian SSH model~\cite{Chang2020,Fossati2023,MunozArboleda2024,YangZhenghao2024}, the non-Hermitian quasicrystal model~\cite{Chen2022,Zhou2023,Zhou20241,Li2024}, the non-Hermitian fermionic models~\cite{Wei2023}, the long-range non-Hermitian SSH model~\cite{Shunlin2024}, and the non-Hermitian Kitaev chain~\cite{Zhou20242}. Apart from the non-Hermitian free-fermion systems, entanglement is studied in specific non-Hermitian interacting systems, such as the non-Hermitian Ising chain~\cite{Bianchini2015,Tista2023,Chao2023}, the non-Hermitian XY chain~\cite{Couvreur2017,LiYue2023,Turkeshi2023,MiaoChuanzheng2024}, the bosonic Hatano-Nelson model~\cite{LuChao2024}, the interacting non-Hermitian Aubry-Andr\'e model~\cite{Qian2024}, and one-dimensional non-Hermitian SSH interacting systems~\cite{Chen2022Wei}. Entanglement dynamics of non Hermitian systems has also been widely studied and exhibits rich phenomena~\cite{Sayyad2021,Dynamics2021,Kawabata2023,Le2023,Orito2023,Li2024,Zhou2024,LiZigeng2024,Soares2024}. The time evolution of entanglement spectrum reflects the topology of Hermitian quench dynamics~\cite{GongUeda2018,ChangPY2018,LuShuangyuan2019,Pastori2020}. For example, the entanglement spectrum has crossings to faithfully reflect both $\mathbb{Z}$ (class BDI) and  $\mathbb{Z}_2$ (class D) topological characterizations~\cite{GongUeda2018}; the entanglement spectrum has mid-gap states forming Dirac cones or rings for 2+1 and 3+1 dimensional post-quench state if the post-quench order parameter has non-vanishing topological invariant~\cite{ChangPY2018}. Similarly, the entanglement spectrum also has crossings and reveals non-Hermitian dynamical topology~\cite{Sayyad2021}. In addition, entanglement entropy is also applied to  describe the time evolution of non-Hermitian systems by tuning quench parameters and non-Hermitian parameters, with entanglement phase transitions being driven by unitary dynamics and non-Hermitian effects~\cite{Dynamics2021,Kawabata2023,Le2023,Orito2023,Li2024,Zhou2024,LiZigeng2024,Soares2024}. For instance, the skin effect induces nonequilibrium quantum phase transitions in entanglement dynamics~\cite{Kawabata2023}, the many-body Hatano-Nelson model exhibits characteristic nonmonotonic time evolution~\cite{Orito2023}, and non-Hermitian Floquet systems reveal rich patterns of entanglement transitions~\cite{Zhou2024}.

Besides entanglement entropy, several other useful entanglement quantities are also borrowed from Hermitian cases to diagnose non-Hermitian physics. It is found that the definition of entanglement quantities and the energy gap are vital  in determining whether the entanglement spectrum can characterize the topological properties of non-Hermitian systems~\cite{Herviou2019,Taberner2022}. The entanglement spectrum can powerfully exhibit information about the topology of the Hamiltonian in line-gap phases when entanglement quantities are defined by biorthogonal ground states.  Apart from these properties, it is reported that crossings in the time evolution of the entanglement eigenvalues can be used to identify the robust topology of non-Hermitian dynamical systems~\cite{Sayyad2021}, and edge entanglement entropy is used to describe different phases of non-Hermitian systems~\cite{Chen2022Wei}. Based on the distribution of the entanglement spectrum, the delocalized and localized phases of the non-Hermitian quasicrystal model can be distinguished~\cite{Chen2022}.  In Hermitian systems, the quantum mutual information between subsystems $A$ and $B$\footnote{It should be noted that, different from Fig.~\ref{bipartition},  these two subsystems, for the purpose of mutual information calculation, do not necessarily cover the whole systems, i.e., $A\bigcup B$ is smaller than the whole system.} is defined as $I(A:B)=S_A+S_B-S_{AB}$, where $S_A$, $S_B$, and $S_{AB}$ are the von Neumann entropy for $A$, $B$, and composite system $A\bigcup B$, respectively. Non-zero mutual information characterizes the correlation between the two subsystems. Undoubtedly, mutual information also plays a role in characterizing non-Hermitian systems. Ref.~\cite{Cheng2024} investigates the quench dynamics of a  non-Hermitian system with a $\mathbb{Z}_2$ gauge field, and it is proposed that the non-Hermitian quantum disentangled liquids exist both in the localized and delocalized phases by distinguishing diverse scaling behaviors of quantum mutual information.

Entanglement in non-Hermitian systems demonstrates a lot of exotic phenomena that never appear in Hermitian systems, triggering a lot of attention. For instance, entanglement entropy exhibits negative values at critical points in the non-Hermitian SSH model, and the entanglement spectrum shows complex values in both the non-Hermitian SSH model and the Chern insulator~\cite{Chang2020}. While the negative central charge $c=-2$ can be explained by non-unitary CFT, mid-gap states are still preserved in the complex entanglement spectrum, aligning with the topology of non-Hermitian systems. Ref.~\cite{LiuSirui2024} introduces a new class of non-Hermitian critical transitions, called scaling-induced exceptional criticality, which show dramatic dips in entanglement entropy scaling, deviating from the usual logarithmic behavior.  Except for these, some entanglement phenomena are similar to Hermitian systems. For example, in a $\mathcal{PT}$-symmetric non-Hermitian phase, entanglement entropy defined by only right eigenstates (or, only left eigenstates) shows  identical behavior to a Hermitian system that connects to the non-Hermitian system by a similarity transformation~\cite{Modak2021}.

In the following, we review some specific models to illustrate the properties of entanglement in non-Hermitian systems.

As a $1$D Hermitian paradigmatic quasicrystal lattice model, the Aubry-Andr\'e-Harper  model is deeply studied for the properties of localization and the critical point. The generalized non-Hermitian quasicrystal models have multiple forms. One of   non-Hermitian quasicrystal models is the model with asymmetric hopping and incommensurate complex potential, written as
$H=\sum_n(J_R c^{\dag}_{n+1}\hat{c}_n+J_L\hat{c}_n^{\dag}\hat{c}_{n+1})+\sum_n\mathcal{V}_n \hat{c}_n^{\dag}\hat{c}_n\,,
 $ where $\hat{c}_n^{\dag}(\hat{c}_n)$ is the creation (annihilation) operator of a spinless fermion at lattice site $n$. $\mathcal{V}_n=V\exp(-2\pi i \alpha n)$ is a site-dependent incommensurate complex potential with irrational number $\alpha$, and the potential strength $V$ is a positive real number. Entanglement can disclose the phase and phase transition of the non-Hermitian quasicrystals~\cite{Chen2022}. On the one hand, the metal-insulator transition point is determined by measuring the entanglement entropy with real-space and momentum-space partitions, as shown in Figs.~1(a) and 1(b) in Ref.~\cite{Chen2022}. On the other hand, the delocalized and the localized phases are characterized by the real-space and momentum-space entanglement spectrum, as shown in Figs.~2(a) and 2(b) in Ref.~\cite{Chen2022}. According to further analyzing for the numerical results in Fig.~2 of Ref.~\cite{Chen2022}, it is proposed that the quasicrystal model with $J_L=0$ has a self-duality between two phases. By performing    Fourier transformations and space inversion,  it is  exactly proved that the self-dual point exists and is also the transition point. The related numerical results are shown in Fig.~3 in Ref.~\cite{Chen2022}. Besides, entanglement can be used to identify the mobility edge in another non-Hermitian quasicrystal model, which is described by the Hamiltonian
 $ H=\sum_n(J_R c^{\dag}_{n+1}\hat{c}_n+J_L\hat{c}_n^{\dag}\hat{c}_{n+1})+\sum_nV(1-ae^{i2\pi \alpha n})^{-1} \hat{c}_n^{\dag}\hat{c}_n\,,
 $ where the special on-site potential induces the mobility edge. At the mobility edge, due to delocalized states suddenly changing to localized states, there are obvious boundary lines in the real-space and momentum-space entanglement spectrum, as shown in Fig.~4 in Ref.~\cite{Chen2022}.

Another interesting model  is the non-Hermitian   SSH model with  $\mathcal{PT}$-symmetry~\cite{Liang2013, Lieu2018, Chang2020}. In momentum space, it is written as
 $ H_k=\bigg(\begin{matrix} iu & \upsilon_k \\ \upsilon_k^{*}  & -i u \end{matrix}\bigg)\,,
  $ where $\upsilon_k=\omega e^{-ik}+\upsilon$ with $u,\upsilon,\omega\in \mathbb{R}$ and  $k$ is the single-particle momentum. Its eigenvalues are $E_{k}=\pm\sqrt{|\upsilon_k|^2-u^2}$. There are three phases with $u\neq 0$, $\mathcal{PT}$-symmetric phases located at the region with $\omega-\upsilon>u$ and  $\omega-\upsilon<-u$, and spontaneously $\mathcal{PT}$-broken phase located at the region with $|\omega-\upsilon|<u$. In the $\mathcal{PT}$-broken phase, the energy spectrum is complex and gapless with two exceptional points. The entanglement entropy at the critical point between the trivial $\mathcal{PT}$-symmetric gapped phase and $\mathcal{PT}$-symmetric broken phase exhibits a logarithmic scaling $S_A(L_A)=-8.81185-0.666\ln[\sin(\pi L_A/L)]$ with central charge $c=-2$, as shown in Fig.~2(a) in Ref.~\cite{Chang2020}, which relates to the $bc$-ghost CFT. The eigenvalues of the correlation matrix still can describe the topological property by showing two mid-gap states in the topological $\mathcal{PT}$-symmetric phase, as shown in Fig.~3(a) in Ref.~\cite{Chang2020}. In this phase, two mid-gap states have imaginary parts, as shown in Fig.~3(b) in Ref.~\cite{Chang2020}. Due to $\xi_n=0.5\pm i I_n$, $\xi_n$ and $1-\xi_n$ are complex conjugate to each other and the imaginary parts do not contribute to entanglement entropy.

The Hamiltonian of a non-Hermitian Chern insulator in momentum space is written as
 $H_{ {k}}=(m+t\cos k_x+t\cos k_y)\sigma_x
+(i\gamma+t\sin k_x)\sigma_y+(t\sin k_y)\sigma_z$,  where $t,m$, and $\gamma$ are real parameters. The topologically non-trivial gapped phases can be characterized by the Chern number. In the topologically non-trivial gapped phases $(t=1.0,m=-1.0,\gamma=0.5)$, the entanglement spectrum with the entanglement cut along the $x$ direction shows the mid-gap states with the imaginary part but entanglement spectrum with the entanglement cut along the $y$ direction doesn't have the imaginary part, as shown in Fig.~4 in Ref.~\cite{Chang2020}.

Subsequently, some researchers focus on negative entanglement entropy in non-Hermitian free-fermion systems by introducing exceptional bound (EB) states~\cite{Lee2022,ZOU2024Deyuan} and topologically protected negative entanglement~\cite{Wen2024}. EB states are robust boundary-induced states and arise from geometric defectiveness. Specifically, they arise at the exceptional gapless points with robust anomalously large or negative occupation probabilities. The anomalously large EB occupancy is encoded in the reduced density matrix and results in negative entanglement entropy. A generalized non-Hermitian SSH model is considered~\cite{Lee2022}
 $H(k)=(\nu-w\cos k)\sigma_x+\gamma_0 \sin k \sigma_y +i(\nu-w)\sigma_z\,.
 $ For the convenience of calculation, they swap $\sigma_y$ and $\sigma_z$, and the Hamiltonian becomes a special form. Considering the long-wavelength limit, the Hamiltonian becomes $H={\small\bigg( \begin{matrix} \gamma_0 k^{\Gamma} & a_0\\ b_0k^B &-\gamma_0 k^{\Gamma}\end{matrix}\bigg)}$, where $B=2,a_0=2(\nu-w),b_0=w/2$. When increasing $\gamma_0$ from 0, the central charge in entanglement entropy
 \begin{equation}
 S\sim(c/3)\ln L
 \end{equation} increases from $-2$ to 1. When increasing $B>2$, the central charge crosses over from $1$ to $-3(B-2)$.

In addition, topologically protected negative entanglement means that the negative entanglement entropy arises when topological edge modes intersect at an exceptional crossing~\cite{Wen2024}.  A  four-band model is taken as an example,
  $H(k,k_y)\!\!=(\cos k_y -\sin k -M)\tau_x \sigma_0+\tau_y(\cos k \sigma_x -\sigma_y+\sin k_y \sigma_z)
+(\sin \alpha \tau_0 +\cos \alpha \tau_x)\sum_{\mu=x,y,z}\sigma_{\mu}+i \delta \tau_y \sigma_0\,,$ where the $\sigma_{\mu}$ and $\tau_{\mu}$ Pauli matrices act in spin and sublattice space, respectively. Non-Hermiticity arises at sublattice hopping asymmetry realized by $i \delta \tau_y \sigma_0$. Consider cylindrical geometry, the topological edge states are bounded along $y$ direction. In the non-trivial Chern insulator case $(\alpha=0, M=3, \delta=2)$, two topological edge modes cross with $\eta(k)=1$. $\eta$ is the generalized Petermann factor that can measure the strength of non-orthogonality between different eigenstates \cite{Zou2023},
\begin{equation}
\eta=\frac{ |\langle \psi^R_{m}|\psi^R_{n}\rangle|^2}{ |\langle \psi^R_{m}|\psi^R_{m}\rangle||\langle \psi^R_{n}|\psi^R_{n}\rangle|}\,, 
\end{equation}
where $0\leq\eta\leq1$. the two eigenstates are mutually orthogonal when $\eta=0$ and the two eigenstates are coalescent when $\eta=1$. The two topological edge modes are parallel, and arise anomalously large EB occupancy that contributes to negative entanglement entropy. The real part of entanglement entropy scales as $\text{Re}S\sim -0.3399\ln L \approx (\frac{1}{3}-\frac{2}{3})\ln L$, where $\frac{1}{3}\ln L$ is attributed to gapless non-exceptional crossing, $-\frac{2}{3}\ln L$ is attributed to gapless exceptional crossing, as shown in Fig.~1 in Ref.~\cite{Wen2024}.

The Fermi surface in non-Hermitian systems becomes complex and the geometry of the Fermi surface has a non-trivial effect on the entanglement entropy~\cite{Guo2021}. It is found that each Fermi point contributes exactly $1/2$ to the coefficient $c$ of the logarithmic correction of entanglement entropy for low-dimensional systems, scaling as  $ 
S=\frac{c}{3}\ln L+O(1)\,,
 $ 
 where $c=\frac{N_f}{2}$ in the thermodynamic limit $L\rightarrow \infty$, $N_f$ is the number of fermi point. Concretely, consider a 1D spinless fermions model with $n$ sublattices in a unit cell,
 $H_{1D}=\sum_{i=1}^L (t_i^L \hat{c}_i^{\dag}\hat{c}_{i+1}+t_i^R \hat{c}_{i+1}^{\dag}\hat{c}_i)
 $ with the hopping constants $t_i^L=t^L, t_i^R=t^R$ for $i=n(l-1)+1$ and $t_i^L=t_i^R=t$ for otherwise, where $l$ is the $l$th unit cell, $t=1,t^L=1+\gamma/2$, and $t^R=1-\gamma/2$. When $n=2$ and half-filling the real part or imaginary part of energy spectra, its number of Fermi points changes as $\gamma$ increases and the system undergoes a Lifshitz phase transition. Correspondingly, the entanglement entropy with real-energy half-filled ground states suddenly increases at $\gamma>\gamma_c$ since the number of Fermi points doubles as shown in Fig.~3(a) in Ref.~\cite{Guo2021}.
By generalizing the 1D model   to 2D, then the model becomes
 $H_{2D}=\sum_{i,j}[(\hat{c}_{2i,j}^{\dag}\hat{c}_{2i+1,j}+\hat{c}_{j,2i}^{\dag}\hat{c}_{j,2i+1}+h.c.)+t^L(\hat{c}_{2i-1,j}^{\dag}\hat{c}_{2i,j}+\hat{c}_{j,2i-1}^{\dag}\hat{c}_{j,2i})+t^R(\hat{c}_{2i,j}^{\dag}\hat{c}_{2i-1,j}+\hat{c}_{j,2i}^{\dag}\hat{c}_{j,2i-1})]\,,
$ where $t^L=1+\frac{\gamma}{2}$ and $t^R=1-\frac{\gamma}{2}$.   When taking $(L_y=4)$ and large $L_y$ with periodic boundary condition in both the $x$ and $y$ directions, entanglement entropy with entanglement cut along $y$ direction is fit as logarithmic form with $L_x$, and the central charge is equal to half the number of Fermi points, as shown in Fig.~6 in Ref.~\cite{Guo2021}.

Faced with the complex spectra and ill-defined ground state in the effective non-Hermitian Hamiltonian as well as the absence of the ground state of open quantum systems, it naturally raises a question of what is the robust signature of non-Hermitian topology in the open many-body systems.  Recent study shows that   the notion of entanglement eigenvalues crossings plays an important role~\cite{Sayyad2021}. Consider a Kitaev chain at $t=0$, the Hamiltonian of the Kitaev chain is described by
 $ H =\!\!\sum_{i=1}^N[-J\hat{c}_i^{\dag}\hat{c}_{i+1}+\Delta \hat{c}_i \hat{c}_{i+1}+H.c.-\mu(\hat{c}_i^{\dag}\hat{c}_i-\frac{1}{2})]\,,
 $ where $J$ is the hopping strength, $\Delta$ is the $p$-wave pairing amplitude and $\mu$ is the chemical potential. Then they abruptly change the chemical potential as well as the system couples with Markovian baths. The corresponding jump operator is $\hat{L}_i=\sqrt{\gamma_1} \hat{c}_i+\sqrt{\gamma_g} \hat{c}_i^{\dag}$, where $\gamma_1$ is the loss rate of particles and $\gamma_g$ is the gain rate of particles, and the Hermitian jump operator is corresponding to balanced gain and loss rates, $\gamma_l=\gamma_g=\gamma$. The pure initial state $\rho_0=|\psi_0\rangle\langle \psi_0|$ experiences coupling with Markovian baths and evolves into a mixed state $\rho(t)=e^{\mathcal{L} t}\rho_0$. The quadratic Liouvillian superoperator $\mathcal{L}$ is constructed by the third quantization based on quadratic Hamiltonian $H$ and linear jump operators $L_i$. On the one hand, the topology of Liouvillian superoperator $\mathcal{L}$ is calculated by non-Hermitian topological band theory. On the other hand, the entanglement eigenvalues of $\rho(t)$ are calculated. The entanglement eigenvalues crossings arise at the topological phase realized by postquench Liouvillian, and the crossing points are located at the points that switch fermion parity of the entanglement ground states, as shown in Fig.~1 in Ref.~\cite{Sayyad2021}. As for the non-Hermitian dissipative systems without quantum jumps, the time-evolved state is represented as $|\psi(t)\rangle=e^{-iH_{\text{eff}} t}|\psi_0\rangle/||e^{-iH_{\text{eff}} t}|\psi_0\rangle||$, and the density matrix is generally defined by only the wave function $|\psi(t)\rangle$ from a measurement perspective~\cite{Herviou2019,Dynamics2021,Kawabata2023}.

\section{Conclusions}\label{sec6}

In summary, we briefly outline the realization of non-Hermitian quantum systems from open systems through the Lindblad master equation, alongside recent studies on entanglement in non-Hermitian free-fermion quantum systems, emphasizing the novel phenomena that arise. Entanglement plays a crucial role in describing the topology and reflecting the non-trivial quantum correlations presenting in both non-Hermitian and open systems. The complex-valued  energy spectra, combined with the biorthogonality and geometric defects of non-Hermitian matrices, give rise to intricate entanglement behaviors. In addition to the non-Hermitian fermion systems, non-Hermitian spin systems, such as the Ising model with imaginary magnetic fields~\cite{Lee1952, Gehlen1991, Wei2012, Peng2015} and the XXZ spin chain with imaginary boundary magnetic fields~\cite{Pasquier1990, Alcaraz1987, Pasquier1990}, are extensively studied. In these systems, researchers analyze entanglement entropy and central charge within the framework of non-unitary CFT~\cite{Bianchini2015, BIANCHINI2015835, Bianchini2016}.

There are many open questions regarding entanglement in non-Hermitian quantum systems and open quantum systems, and we conclude our discussion by highlighting several possible future directions. From the quantum information perspective, how should we understand the negative entanglement entropy observed in non-Hermitian quantum systems? How do non-orthogonal bases and complex energy spectra intrinsically influence the complex entanglement entropy and entanglement spectrum? Additionally, when considering the presence of the non-Hermitian skin effect in non-Hermitian topological systems, the conventional bulk-boundary correspondence becomes in-valid~\cite{Yao2018,Kunst2018}. In contrast, the entanglement spectrum offers an alternative approach to study this correspondence in Hermitian systems~\cite{Li2008,Fidkowski2010}. Following the framework established for Hermitian systems, Refs.~\cite{Herviou2019, Chang2020} suggest to utilize the entanglement spectrum to investigate the bulk-boundary correspondence in non-Hermitian systems. Due to the ambiguity in the definition of density matrix, a systematic study of the entanglement spectrum is still lacking in non-Hermitian systems. We believe that the entanglement spectrum could provide valuable insights into the bulk-boundary correspondence in these systems. In conclusion, these questions remain open and warrant further investigation.

Given a non-Hermitian quantum system, are there experimental platforms capable of measuring the entanglement behaviors of non-Hermitian free-fermion systems or spin systems? Experimental simulations could provide valuable insights into questions surrounding the definition of entanglement entropy and address unphysical behaviors such as negative entanglement entropy and complex entanglement spectra. Notably, the area law of quantum mutual information is verified using an ultracold atom simulator~\cite{TajikKukuljan2023}, and the experimental measurement of entanglement properties in Hermitian free-fermion systems is reported in phononic systems~\cite{Lin2024}. Similar efforts may also be made for measuring entanglement in both Hermitian and non-Hermitian systems in square, fractal~\cite{ZhouYaofractal2023} and hyperbolic lattices~\cite{zhouyaohyperbolic} since gain and loss are natural in these apparatuses.

Apart from free-fermion systems, entanglement encodes the number of Nambu-Goldstone modes~\cite{Metlitski2011} in phases of spontaneously broken continuous symmetry. One can study entanglement scaling in more exotic spontaneous symmetry-breaking phases, namely, fractonic superfluids, in which the symmetry is associated with the conservation of both charges and dipoles, or much higher moments~\cite{fs_1,fs_2,fs_3,fs_4,fs_5,fs_review}. Entanglement study of such boson systems with / without environment may be of great interests.

	\acknowledgements
This work was in part supported by National Natural Science Foundation of China (NSFC) Grant No. 12074438. The calculations reported were performed on resources provided by the Guangdong Provincial Key Laboratory of Magnetoelectric Physics and Devices, No. 2022B1212010008.


%

\end{document}